\documentclass[%
apl,aip,showpacs,numerical,amsmath,amssymb,floatfix,
reprint,natbib%
]{revtex4-1}

\usepackage{graphicx}
\usepackage{dcolumn}
\usepackage{bm}
\usepackage[english]{babel}
\usepackage[utf8]{inputenc}
\usepackage[separate-uncertainty = true,multi-part-units=single]{siunitx}
\usepackage{xcolor}

\draft
\begin{document}

\preprint{DS001}

\title{Magnetoelasticity of Co$_{25}$Fe$_{75}$ thin films} 

\author{Daniel Schwienbacher}
\email[]{daniel.schwienbacher@wmi.badw.de}
\affiliation{Walther-Meissner-Institut, Bayerische Akademie der Wissenschaften, Walther Meissner Str.8 85748 Garching}
\affiliation{Technische Universität München, James-Franck-Str. 1, 85748 Garching, Germany}
\affiliation{Munich Center for Quantum Science and Technology (MCQST), Schellingstraße 4, 80799 München, Germany}

\author{Matthias Pernpeintner}
\altaffiliation{}
\affiliation{Walther-Meissner-Institut, Bayerische Akademie der Wissenschaften, Walther Meissner Str.8 85748 Garching}
\affiliation{Technische Universität München, James-Franck-Str. 1, 85748 Garching, Germany}

\author{Lukas Liensberger}
\altaffiliation{}
\affiliation{Walther-Meissner-Institut, Bayerische Akademie der Wissenschaften, Walther Meissner Str.8 85748 Garching}
\affiliation{Technische Universität München, James-Franck-Str. 1, 85748 Garching, Germany}

\author{Eric R.J.Edwards}
\altaffiliation{}
\affiliation{National Institute of Standards and Technology, Boulder, Colorado 80305, USA}

\author{Hans T. Nembach}
\altaffiliation{}
\affiliation{National Institute of Standards and Technology, Boulder, Colorado 80305, USA}

\author{Justin M. Shaw}
\altaffiliation{}
\affiliation{National Institute of Standards and Technology, Boulder, Colorado 80305, USA}

\author{Mathias Weiler}
\altaffiliation{}
\affiliation{Walther-Meissner-Institut, Bayerische Akademie der Wissenschaften, Walther Meissner Str.8 85748 Garching}
\affiliation{Technische Universität München, James-Franck-Str. 1, 85748 Garching, Germany}

\author{Rudolf Gross}
\altaffiliation{}
\affiliation{Walther-Meissner-Institut, Bayerische Akademie der Wissenschaften, Walther Meissner Str.8 85748 Garching}
\affiliation{Technische Universität München, James-Franck-Str. 1, 85748 Garching, Germany}
\affiliation{Munich Center for Quantum Science and Technology (MCQST), Schellingstraße 4, 80799 München, Germany}

\author{Hans Huebl}
\altaffiliation{}
\email[]{huebl@wmi.badw.de}
\affiliation{Walther-Meissner-Institut, Bayerische Akademie der Wissenschaften, Walther Meissner Str.8 85748 Garching}
\affiliation{Technische Universität München, James-Franck-Str. 1, 85748 Garching, Germany}
\affiliation{Munich Center for Quantum Science and Technology (MCQST), Schellingstraße 4, 80799 München, Germany}

\date{\today}

\begin{abstract}
	We investigate the magnetoelastic properties of $\mathrm{Co_{25}}\mathrm{Fe_{75}}$ and $\mathrm{Co_{10}}\mathrm{Fe_{90}}$ thin films by measuring the mechanical properties of a doubly clamped string resonator covered with multi-layer stacks containing these films. For the magnetostrictive constants we find $\lambda_{\mathrm{Co_{25}}\mathrm{Fe_{75}}}=(-20.68\pm0.25)\times10^{-6}$ and $\lambda_{\mathrm{Co_{10}}\mathrm{Fe_{90}}}=(-9.80\pm0.12)\times10^{-6}$ at room temperature. In stark contrast to the positive magnetostriction previously found in bulk CoFe crystals. $\mathrm{Co_{25}}\mathrm{Fe_{75}}$ thin films unite low damping and sizable magnetostriction and are thus a prime candidate for micromechanical magnonic applications, such as sensors and hybrid phonon-magnon systems.
\end{abstract}


\maketitle 

\begin{figure}
 	\includegraphics[scale=1]{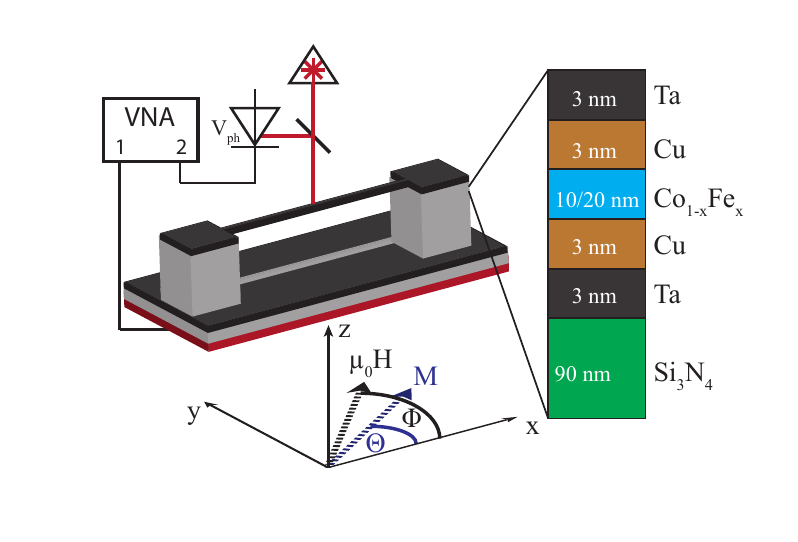}
 	\caption{Schematic of doubly clamped $\mathrm{Si_{3}}\mathrm{N_{4}}$\,nano-string covered with a CoFe layer stack on top (black) and interferometric readout setup. The string is supported by posts etched from the silicon substrate (grey). The whole sample is mounted on a piezo-actuator (red). $\Phi$ is the angle between the external magnetic field and the \textit{x}-direction (along the string), $\Theta$ denotes the angle between the \textit{x}-direction and the magnetization direction in the CoFe film. The layer stack with Ta($\SI{3}{\nm}$) and Cu($\SI{3}{nm}$) seed and capping layers is the same as used in Ref.\,\citenum{Schoen2016}. The CoFe layer thickness varied for different alloy ratios. To extract the resonance frequency of the oop mechanical motion of the string, the amplitude of a reflected laser beam is measured with a photo-diode and analyzed with a VNA.
 	}
 	\label{fig:setup}
 \end{figure}

Magnetic alloys are an extremely well studied material group due to their importance for applications in magnetic information storage. While properties such as the saturation magnetization and magnetic anisotropy play key roles for the static configuration and stability of the magnetization state, material parameters related to magnetization control (beyond such enacted by static magnetic fields) are also of interest. Apart from current-induced magnetization switching\cite{Mangin2006,Krause2007,Yang2008,Miron2011,Liu2012}, techniques based on magnetostriction constitute a complementary way to control the magnetization direction. Here, the elastic deformation of the material generates a strain-induced anisotropy term which can be used to reorient the magnetization. Static control \cite{Weiler2009,Brandlmaier2011,Spaldin2005,Martin2012} as well as the excitation of magnetization dynamics \cite{Dreher2012,Weiler2011,Gowtham2015,Chang2018} already has been demonstrated. Moreover, the reciprocal effect is used in sensing applications based on magnetoelastics \cite{Grimes2011}.\\
Cobalt iron alloys recently regained interest as an electrically conducting ferromagnetic material with ultra-low damping \cite{Schoen2016,Schoen2017,Flacke2019}. Damping in thin film Co$_{25}$Fe$_{75}$ was found to be as low as in thin film yttrium iron garnet \cite{Collet2017}. Since applications in spin electronics are usually based on thin films, quantification of the magnetoelastic properties of thin film Co$_{25}$Fe$_{75}$ is required. In previous studies \cite{Hall1959,Hunter2011,Quandt1999,Quandt2000,Quandt1998,Ludwig2002,Chopra2005,Johnson2004,Cooke2001,Cooke2000,Zuberek2000,Hung2000,Cheng2018} only bulk materials have been studied.\\
In this article, we investigate the magnetostrictive properties of $\mathrm{Co_{25}}\mathrm{Fe_{75}}\,\mathrm{and}\,\mathrm{Co_{10}}\mathrm{Fe_{90}}$
thin films. The films were grown using the same recipe as the ultra-low damping material of Ref.\,\citenum{Schoen2016}. In our study we employ magnetostriction measurements based on nano-strings as reported in Ref.\citenum{Pernpeintner2016}.
The paper is organized as follows: First we briefly sketch the physics of the nano-strings and how it is influenced by the magnetoelastic properties of the $\mathrm{Co_{x}}\mathrm{Fe_{1-x}}$ thin films deposited on them. Then we give a short description of sample fabrication, and the experimental setup used to characterize them. We then provide an in depth data analysis and summarize our findings. \\
To access the magnetostrictive properties of thin film $\mathrm{Co_{x}}\mathrm{Fe_{1-x}}$, we deposit the ultra-low magnetization damping layer stacks reported in Ref.\citenum{Schoen2016} onto a freely suspended silicon nitride string (c.f. Fig.\,\ref{fig:setup}). The resonance frequency of this multilayer string scales approximately with $1/L\sqrt{\sigma_\mathrm{eff}/\rho_\mathrm{eff}}$, where $L$ is the length of the string, $\sigma_\mathrm{eff}$ is the effective stress along the string, and $\rho_\mathrm{eff}$ is the effective mass density of the whole layer stack. $\sigma_\mathrm{eff}$ is directly related to the the static stress $\sigma_\mathrm{0}$ in the system. Moreover, when we measure the resonance frequency as a function of the magnetization direction, we expect a modulation of the resonance frequency because the magnetoelastic interaction changes the stress in the sample depending on the magnetization direction. In more detail, the resonance frequency of a highly tensile stressed, doubly clamped nano-string, also depends on material parameters, like the Young's modulus $E$, and size dependent parameters like the string moment of inertia $I$ and its cross-section $A=wt$, where $w$ is the strings width and $t$ its thickness. A nano-string can be treated as highly tensile stressed, if the static stress $\sigma_0$ is the dominant parameter ($(4 E w^2 t^4 \rho  \Omega_0^2) /12 \ll \sigma_0^2 (wt)^2$). The magnetization direction dependent resonance frequency of the string is given by \cite{Timoshenko2008,Verbridge2006}
\begin{equation}
 \Omega_\mathrm{0}=\frac{\left( \sigma_0+\sigma_1 \cos\Theta^2\right)\pi\sqrt{\rho_\mathrm{eff}^{-1}}}{L\sqrt{\sigma_0+\sigma_1 \cos\Theta^2}-2\sqrt{Et^2/12}}.
 \label{eq:omega}
\end{equation}
This equation includes geometry sensitive bending effects to first order. The magnetization orientation with respect to the long axis of the string is denoted by $\Theta$, and $\sigma_1$ determines the change in stress along the \textit{x}-direction. Note, however, that $\Theta$ is not directly accessible in our experiment. Our data are rather recorded as a function of the applied magnetic field direction, which is given by the angle $\Phi$ (see Fig.\,\ref{fig:setup}). To relate $\Phi$ to $\Theta$ we calculate the magnetization direction $\Theta$ for a given external magnetic field $\mathbf{H}$ by using a free energy minimization approach. For a uniaxial anisotropy along $x$ we obtain:
\begin{equation}
\Theta(\Phi)=\Phi - \frac{K \sin(2\Phi)}{-M_\textrm{S} \mu_0 H + 2 K \cos(2\Phi)},
 \label{eq:phiofthetha}
\end{equation}
 with the saturation magnetization $M_\textrm{S}$ and the uniaxial anisotropy constant $K$ \cite{GrossMarx}.\\
 With the relation \eqref{eq:phiofthetha} we can translate the measured $\Omega_0(\Phi)$ dependence into an $\Omega_0(\Theta)$ dependence, which is fitted by Eq.,\eqref{eq:omega} to derive the stress component $\sigma_1$. The derived value of $\sigma_1$ finally allows us to determine the magnetostrictive constant\cite{Pernpeintner2016}:
\begin{equation}
 \lambda_\parallel = \frac{\sigma_1 t}{t_\mathrm{CoFe} E_\mathrm{CoFe}}.
 \label{eq:lambda}
\end{equation} 
 Note that due to the specific geometry of the string, we can access only the parallel part ($\lambda_\parallel \parallel \textit{x}$) of the magnetostrictive constant, because only stress variations in the \textit{x}-direction change the string's resonance frequency. The quantity $\lambda_\parallel$ used here is equivalent to the quantity $\lambda_\mathrm{S}$ commonly used for polycrystalline material in the literature\cite{Chikazumi}.
 From this magnetostrictive constant we can calculate the magnetoelastic constant $b_1$\cite{Dreher2012,Chikazumi}
 \begin{equation}
 b_1=-\frac{3\lambda_\parallel G}{M_\mathrm{S}},
 \label{eq:b1}
 \end{equation}
 with shear modulus $G$.\\
\label{SS}
For the fabrication of the freely suspended $\mathrm{Co_{x}}\mathrm{Fe_{1-x}}$ layers on top of silicon nitride string resonators, we start with a single crystalline silicon wafer, which is commercially coated with a $t_\mathrm{SiN}=\SI{90}{\nm}$ thick, highly tensile-stressed, low pressure chemical vapor deposition (LPCVD) grown $\mathrm{Si_{3}}\mathrm{N_{4}}$(SiN) film. We define the geometry of the strings by defining a metal etch mask using electron beam lithography, electron beam evaporation of aluminum, and a lift-off process. The pattern is transferred to the silicon nitride using an anisotropic reactive ion etching (RIE) process to define the SiN strings. Subsequently, a second isotropic RIE process is used to remove the Si substrate below the strings to release them and enable mechanical in-plane(ip) and out-of-plane (oop) motion. The Al etching mask is removed afterwards. The resulting unloaded SiN strings show typical $Q$-factors of about $150,000$ for the ip and oop fundamental modes. As the last fabrication step, a Ta/Cu/CoFe/Cu/Ta layer stack (as shown in Fig. \ref{fig:setup}) was deposited on top of the strings by magnetron sputtering. Thus, the CoFe stack covers the strings as well as the surrounding substrate. However, we ensure that there is no mechanical contact between the top- or string layer and the substrate level. We investigate two sets of strings with two different CoFe alloys: The reported ultra-low damping $\mathrm{Co_{25}}\mathrm{Fe_{75}}$, and $\mathrm{Co_{10}}\mathrm{Fe_{90}}$ as an alloy with larger damping for comparison. For each Co-Fe alloy, we investigate the mechanical response for different string lengths $L$ (\SI{25}{\micro\meter}, \SI{35}{\micro\meter} and \SI{50}{\micro\meter}) and string widths $w$ (\SI{150}{\nano\meter} and \SI{200}{\nano\meter}). We suspect that during the sputtering process material was also deposited on the sides of the string, creating an overhang.\\
\label{exp}
To measure the resonance frequencies of the strings, we use a free space optical interferometer similar to the setup in Ref.\citenum{Pernpeintner2016} (see Fig.\,\ref{fig:setup}). A laser beam with wavelength $\SI{633}{\nm}$ is focused on the string and interferometry is used to measure the amplitude of the nanostring oop motion. To excite the string's oop mode at its resonance frequency, the entire sample is placed on a piezo-actuator (Fig.\,\ref{fig:setup}, red layer). The resonance frequency of the string is obtained by measuring the output voltage of a photo-diode while sweeping the drive frequency using a vector network analyzer (VNA). The drive voltage is chosen small enough to keep the piezo actuator as well as the strings in their respective linear regimes. The sample holder is mounted on a $xyz$ piezo stage to allow positioning and focusing of the laser spot on an individual string. The interferometer is operated at room temperature. The sample stage is placed in vacuum ($p < \SI{0.01}{\pascal}$) to prevent air damping. To control the magnetization, and in particular the magnetization orientation of the Co-Fe on the string, the sample is positioned between the pole pieces of an electromagnet. The applied field direction is varied by rotating the electromagnet whereas the sample position and orientation remain fixed.\\
\begin{figure}
	\includegraphics[scale=1]{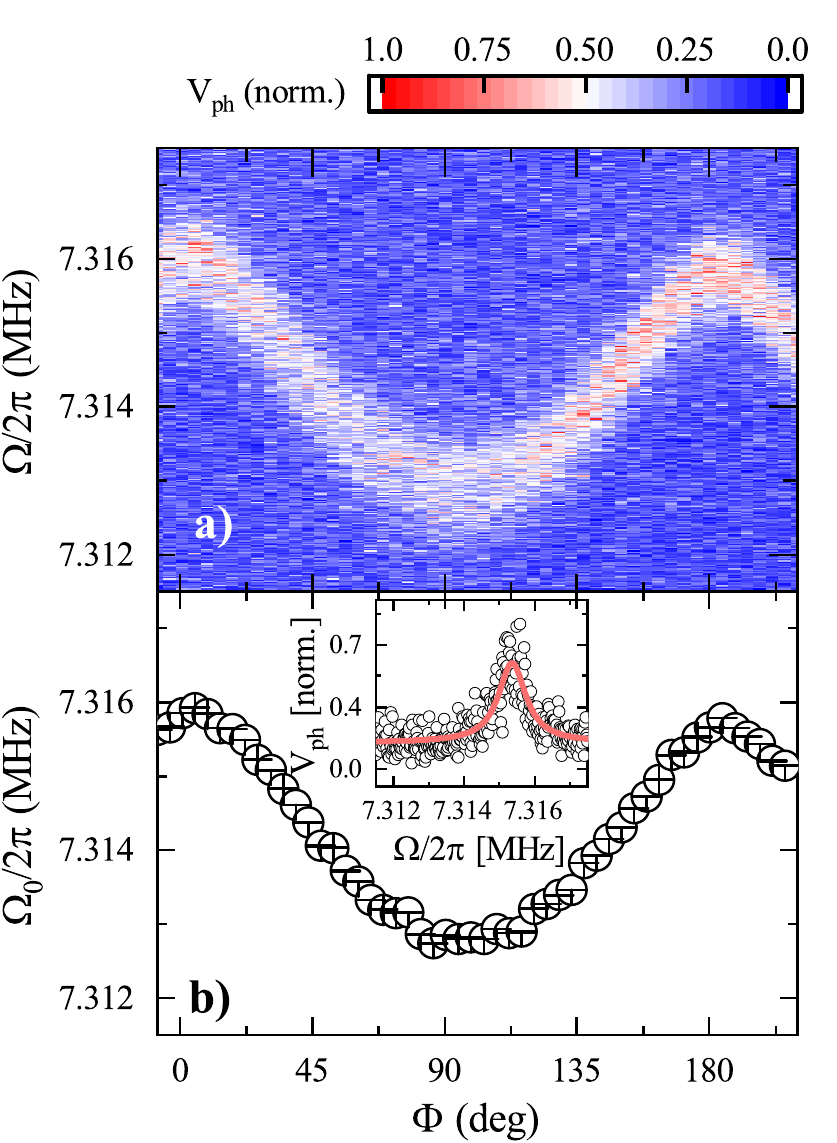}
	\caption{Mechanical response of the fundamental mode of a $\SI{25}{\um}$ long nano-string as a function of external field direction $\Phi$ at $\mu_0 H=\SI{950}{\milli\tesla}$.
		a) Shows the frequency dependent photo-voltage as a function of external magnetic field direction and drive frequency, this is a direct measure for the mechanical amplitude of the string.
		b) Shows the extracted resonance frequencies at specific field directions. The inset in b) shows a slice from a) at $\Phi=\ang{153}$ and the fit to a Lorentzian line shape (red line) used to extract the resonance frequency. Error bars are fit errors.}
	\label{fig:plomerged}
\end{figure}
Figure \ref{fig:plomerged}a) shows a color-encoded plot of the mechanical response function as a function of actuation frequency and applied magnetic field direction. Red highlights large oop mechanical displacement, while blue indicates no visible motion. The raw data is measured for a constant actuation amplitude and a fixed magnitude of the magnetic field $\mu_0 H= \SI{950}{\milli \tesla}$. The resonance frequency of the string is $180^{\circ}$ periodic with respect to the external magnetic field direction.  A cut of this dataset at $\Phi = 153^{\circ}$ is displayed in the inset of Fig.\,\ref{fig:plomerged} b), showing the mechanical response as function of the drive frequency. 
To extract the resonance frequency, we fit a Lorentzian line shape to the data for each measured angle $\Phi$. From this fit we find a linewidth (full-width at half-maximum) of $\SI{900}{\hertz}$ corresponding to a $Q$-factor of the string of about $8000$.
This $Q$-factor is significantly smaller than that of a pure SiN string and can mainly be attributed to the added metal layer stack. The stack increases the overall mass of the string, and thereby its effective density, which lowers the resonance frequency (see Eq.\,\eqref{eq:omega}). Moreover, adding a metal component is known to change the mechanical damping of nano-strings \cite{Seitner2014,Hoehne2010}. \\
Figure \ref{fig:plomerged}b shows the evolution of the resonance frequency as a function of $\Phi$.
\begin{figure}
	\includegraphics[scale=1]{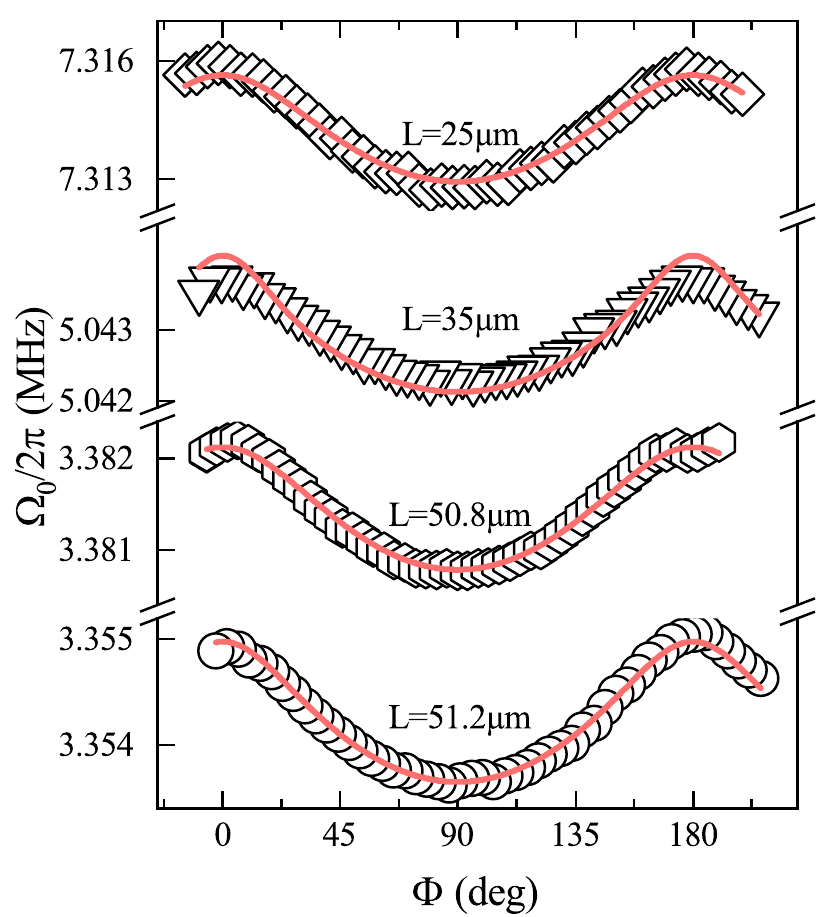}
	\caption{Global fit to magnetization direction dependent resonance frequencies of strings with different lengths covered with the $\mathrm{Co_{25}}\mathrm{Fe_{75}}$ stack. The resonance frequencies of the strings with a length of $\SI{25}{\um}$\,(diamonds), $\SI{35}{\um}\,$(triangles), $\SI{50.8}{\um}\,$(hexagons) and $\SI{51.2}{\um}$\,(circles) length were globally fit using Eqs.\,\eqref{eq:omega} and \eqref{eq:phiofthetha} (red lines). Fit errors are within the size of the data symbols.}
	\label{fig:globalfit25}
\end{figure} 
For comparison, we have measured a set of strings with different lengths and widths. To analyze our data, we use a global fit routine employing Eqs.\,\eqref{eq:omega} and \eqref{eq:phiofthetha}. The fit uses the data of all strings for each CoFe composition as input parameter. In addition, we use the thickness  $t=\SI{112}{\nano\meter}$ of the nano-string and effective density $\rho_\mathrm{eff}=\SI{4350}{\kilogram/\meter^3}$ of each string as fixed parameters, as both are known fabrication parameters. The thickness of the metal stack was determined by calibrating the deposition rates using x-ray reflectometry. The density was calculated by using the weighted average of the single material bulk densities \cite{Cardarelli2018}. 
Figure \ref{fig:globalfit25} shows the fit of $\Omega_0$ for the $\mathrm{Co}_{25}\mathrm{Fe}_{75}$ compound for strings of different lengths. Here, the pre-stress $\sigma_0$, the magnetically induced stress $\sigma_1$, and the Young's modulus $E$ of the sample were set as global fit parameters. For the fit we used fixed values for the length $L$ of the strings with $\SI{25}{\micro\meter}$ and $\SI{35}{\micro\meter}$. The string lengths of the two nominally $\SI{50}{\micro\meter}$ long strings are free fit parameters. This allows to account for small variations in the frequencies of the two nominally identical strings, which otherwise should have the exactly the same frequency. The fitted lengths are $\SI{51.2}{\micro\meter}$ and $\SI{50.8}{\micro\meter}$ in good agreement with the design value of $\SI{50}{\micro\meter}$. The uniaxial anisotropy constant $K$ is a free fit parameter for each string as it might differ from string to string. 
As shown in Fig \ref{fig:globalfit25}, we find good agreement between the global fit and the data using $\sigma_1 = \SI{-386\pm5}{\kilo\pascal}$, $\sigma_0 =\SI{458.7\pm 0.1}{\mega\pascal}$ and $E=\SI{857.7\pm0.2}{\giga\pascal}$. The extracted pre-stress is reduced compared to the pre-stress in a SiN string without any metal on top. This can be attributed to a compressive stress in the layer stack of $\Delta\sigma_0 \approx\SI{270}{\mega\pascal}$. The sputtering process may change the pre-stress of the composite string. Even though the sputtering process is carried out at room temperature, the temperature of the nanostring is expected to increase significantly due to the poor thermal coupling of the string to the substrate. Thus the metal stack is deposited at a temperature well above room temperature. Cooling down the string coated by the metal stack after deposition then results in a partial compensation of the pre-stress due to different thermal expansion coefficients of SiN and the metal stack. A temperature increase of about \SI{300}{\kelvin} could explain the observed change of pre-stress. Also the extracted Young's modulus is larger then expected from the Young's moduli of the individual materials \cite{Cardarelli2018}. Using \eqref{eq:lambda} and \eqref{eq:b1} in combination with the known sample parameters and the Co-Fe Young's modulus we obtain a $\lambda_\parallel=(-20.68\pm0.25)\times10^{-6}$ and $b_1=\SI{2.62\pm0.05}{\tesla}$ for the $\mathrm{Co}_{25}\mathrm{Fe}_{75}$. We obtain these values when considering $t_\textrm{CoFe}=\SI{10}{\nano\meter}$,  $t=\SI{112}{\nano\meter}$,  $E_\textrm{CoFe}=\SI{208}{\giga\pascal}$\cite{Cardarelli2018}, $M_\textrm{S}=\SI{1.904}{\mega\ampere\per\meter}$ \cite{Schoen2017} as well as the shear modulus $G=\SI{81.7}{\giga\pascal}$\cite{Cardarelli2018}.\\
In addition, the measured data allow to access magnetic anisotropy parameters. We find an anisotropy $2K/M_\mathrm{S}\approx\SI{300}{\milli\tesla}$ with an easy axis pointing along the \textit{y}-direction of the string. Note that because we have access only to in-plane measurements, we can calculate only projections of an anisotropy to the \textit{x-y}-plane of the sample. Combined with the calculated shape anisotropy $B_\textrm{shape}\approx\SI{100}{\milli\tesla}$\cite{GrossMarx}, with an easy axis along the \textit{x}-direction of the string, the total anisotropy field in the sample ads up to $B_\textrm{aniso}\approx\SI{400}{\milli\tesla}$. The compressive stress in the metal $\Delta\sigma_0$ leads to a magnetoelastic anisotropy of $B_\mathrm{magel}\approx \SI{4}{\milli\tesla}$ \cite{Chikazumi}. Unfortunately, we cannot identify the origin of the anisotropy. However we speculate that the overhanging material at the edges of the string might result in a preferential orientation of the magnetization direction perpendicular to the string.\\
\begin{figure}
	\includegraphics[scale=1]{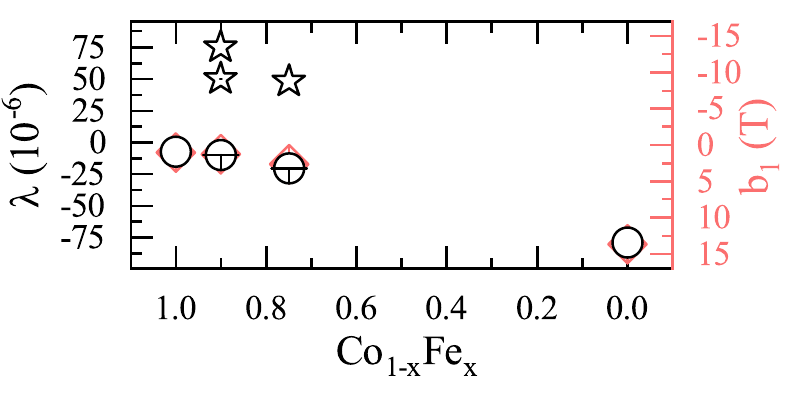}
	\caption{Magnetostrictive and magnetoelastic constants for the two $\mathrm{Co}_{1-x}\mathrm{Fe}_x$ alloys and pure metals (Co\cite{Pernpeintner2016},Fe\cite{Cardarelli2018}) for reference.
		Circles show the magnetostrictive contant ($\lambda_\parallel$) on the left scale, while diamonds (red) depict the corresponding magnetoelastic constant ($b_1$) on the right scale. The star shaped data points correspond to literature values from Refs.\,\citenum{Hunter2011,Hall1959}. Error bars are fit errors.
	}
	\label{fig:lambdaandb1}
\end{figure}
To set these results in context, we plot the extracted values of $\lambda_\parallel$ and $b_1$ for the two measured thin film CoFe alloys ($\mathrm{Co}_{25}\mathrm{Fe}_{75}$ and $\mathrm{Co}_{10}\mathrm{Fe}_{90}$) as well as the values for thin-film Co\cite{Pernpeintner2016} and bulk Fe\cite{Cardarelli2018} in Fig.\,\ref{fig:lambdaandb1}.
The ultra-low damping material investigated in this work seems to follow the simple trend of an interpolating magnetostrictive constant connecting the bulk values. Since the values for the saturation magnetization and shear modulus are similar for Co and Fe the $b_1$ is approximately linearly proportional to $\lambda$. Nevertheless, Fig.\,\ref{fig:lambdaandb1} also shows the data Hunter et al.\cite{Hunter2011} obtained using a cantilever displacement method on various \SI{500}{\nano\meter} thick  $\mathrm{Co}_{x}\mathrm{Fe}_{1-x}$ films. Their data show an entirely different behavior, most importantly an opposing sign of $\lambda_\mathrm{S}\approx 50\times10^{-6}$.
Even earlier experiments by Hall\cite{Hall1959} extrapolated an in-plane magnetostrictive constant of $\lambda_\mathrm{100}\approx 75\times10^{-6}$ for  $\mathrm{Co}_{25}\mathrm{Fe}_{75}$  and of $\lambda_\mathrm{100}\approx 48\times10^{-6}$ for $\mathrm{Co}_{10}\mathrm{Fe}_{90}$ for bulk crystal discs.
We note, however, that the seed layer and the sputtering conditions are crucial for the realization of ultra-low damping material \cite{Edwards2019} and thus rationalize that the magnetoelastic properties can be significantly affected. To ensure that the low-damping behavior of the Co-Fe is still present when changing the substrate from Si\cite{Schoen2016} to SiN used in this paper, we performed ferromagnetic resonance (FMR) experiments on unpatterned CoFe-stacks on SiN samples and find a Gilbert damping of $\alpha =4.2 \pm 0.2\times 10^{-3}$ for a $\SI{20}{\nano\meter}$ thick $\mathrm{Co}_{10}\mathrm{Fe}_{90}$ film which is in agreement with the values from Schoen et al.\cite{Schoen2016}.\\
In this article, we extract the magnetostrictive constants of two low magnetic damping Co-Fe alloys grown within a layer stack\cite{Schoen2016}. To get a quantitative value for the magnetostriction we use a magnetization direction dependent resonance frequency measurement of a nanostring\cite{Pernpeintner2016}, which is covered with the magnetostrictive layer stack. This method allows the investigation of the magnetostrictive and elastic properties of thin film magnetic layers, even with small sample volumes and high aspect ratios, which both are requisites for future technical applications of spintronic devices including sensing applications. 
We extract a magnetostrictive constant of $\lambda_\parallel=(-20.72\pm0.33)\times10^{-6}$ which corresponds to a magnetoelastic constant of $b_1=\SI{2.62\pm0.05}{\tesla}$ for the ultra low damping  $\mathrm{Co}_{25}\mathrm{Fe}_{75}$ compound, as well as $\lambda_\parallel=(-9.8\pm0.12)\times10^{-6}$ and $b_1=\SI{1.3\pm0.02}{\tesla}$ for the $\mathrm{Co_{10}}\mathrm{Fe_{90}}$ compound. This shows that the magnetoelastic properties of the two investigated alloys have the same order of magnitude as the constituent materials but differ significantly between the low-damping and normal damping case. Thus CoFe and in particular the ultra-low damping compound $\mathrm{Co}_{25}\mathrm{Fe}_{75}$ shows a sizeable magnetoelastic constant and hereby makes an ideal candidate for sensing and magnetization dynamic applications which rely on low damping materials.\\
See supplementary material for the derivation of Eq.\,\eqref{eq:omega} and the reference broadband ferromagnetic resonance measurements of thin film CoFe grown on SiN substrate.\\
Funded by the Deutsche Forschungsgemeinschaft (DFG, German Research Foundation) under Germany’s Excellence Strategy – EXC-2111 – 390814868 and project WE5386/4-1.


\begin{thebibliography}{40}%
	\makeatletter
	\providecommand \@ifxundefined [1]{%
		\@ifx{#1\undefined}
	}%
	\providecommand \@ifnum [1]{%
		\ifnum #1\expandafter \@firstoftwo
		\else \expandafter \@secondoftwo
		\fi
	}%
	\providecommand \@ifx [1]{%
		\ifx #1\expandafter \@firstoftwo
		\else \expandafter \@secondoftwo
		\fi
	}%
	\providecommand \natexlab [1]{#1}%
	\providecommand \enquote  [1]{``#1''}%
	\providecommand \bibnamefont  [1]{#1}%
	\providecommand \bibfnamefont [1]{#1}%
	\providecommand \citenamefont [1]{#1}%
	\providecommand \href@noop [0]{\@secondoftwo}%
	\providecommand \href [0]{\begingroup \@sanitize@url \@href}%
	\providecommand \@href[1]{\@@startlink{#1}\@@href}%
	\providecommand \@@href[1]{\endgroup#1\@@endlink}%
	\providecommand \@sanitize@url [0]{\catcode `\\12\catcode `\$12\catcode
		`\&12\catcode `\#12\catcode `\^12\catcode `\_12\catcode `\%12\relax}%
	\providecommand \@@startlink[1]{}%
	\providecommand \@@endlink[0]{}%
	\providecommand \url  [0]{\begingroup\@sanitize@url \@url }%
	\providecommand \@url [1]{\endgroup\@href {#1}{\urlprefix }}%
	\providecommand \urlprefix  [0]{URL }%
	\providecommand \Eprint [0]{\href }%
	\providecommand \doibase [0]{http://dx.doi.org/}%
	\providecommand \selectlanguage [0]{\@gobble}%
	\providecommand \bibinfo  [0]{\@secondoftwo}%
	\providecommand \bibfield  [0]{\@secondoftwo}%
	\providecommand \translation [1]{[#1]}%
	\providecommand \BibitemOpen [0]{}%
	\providecommand \bibitemStop [0]{}%
	\providecommand \bibitemNoStop [0]{.\EOS\space}%
	\providecommand \EOS [0]{\spacefactor3000\relax}%
	\providecommand \BibitemShut  [1]{\csname bibitem#1\endcsname}%
	\let\auto@bib@innerbib\@empty
	\bibitem [{\citenamefont {Mangin}\ \emph {et~al.}(2006)\citenamefont {Mangin},
		\citenamefont {Ravelosona}, \citenamefont {Katine}, \citenamefont {Carey},
		\citenamefont {Terris},\ and\ \citenamefont {Fullerton}}]{Mangin2006}%
	\BibitemOpen
	\bibfield  {author} {\bibinfo {author} {\bibfnamefont {S.}~\bibnamefont
			{Mangin}}, \bibinfo {author} {\bibfnamefont {D.}~\bibnamefont {Ravelosona}},
		\bibinfo {author} {\bibfnamefont {J.~A.}\ \bibnamefont {Katine}}, \bibinfo
		{author} {\bibfnamefont {M.~J.}\ \bibnamefont {Carey}}, \bibinfo {author}
		{\bibfnamefont {B.~D.}\ \bibnamefont {Terris}}, \ and\ \bibinfo {author}
		{\bibfnamefont {E.~E.}\ \bibnamefont {Fullerton}},\ }\href {\doibase
		10.1038/nmat1595} {\bibfield  {journal} {\bibinfo  {journal} {Nature
				Materials}\ }\textbf {\bibinfo {volume} {5}},\ \bibinfo {pages} {210}
		(\bibinfo {year} {2006})}\BibitemShut {NoStop}%
	\bibitem [{\citenamefont {Krause}\ \emph {et~al.}(2007)\citenamefont {Krause},
		\citenamefont {Berbil-Bautista}, \citenamefont {Herzog}, \citenamefont
		{Bode},\ and\ \citenamefont {Wiesendanger}}]{Krause2007}%
	\BibitemOpen
	\bibfield  {author} {\bibinfo {author} {\bibfnamefont {S.}~\bibnamefont
			{Krause}}, \bibinfo {author} {\bibfnamefont {L.}~\bibnamefont
			{Berbil-Bautista}}, \bibinfo {author} {\bibfnamefont {G.}~\bibnamefont
			{Herzog}}, \bibinfo {author} {\bibfnamefont {M.}~\bibnamefont {Bode}}, \ and\
		\bibinfo {author} {\bibfnamefont {R.}~\bibnamefont {Wiesendanger}},\ }\href
	{\doibase 10.1126/science.1145336} {\bibfield  {journal} {\bibinfo  {journal}
			{Science}\ }\textbf {\bibinfo {volume} {317}},\ \bibinfo {pages} {1537}
		(\bibinfo {year} {2007})}\BibitemShut {NoStop}%
	\bibitem [{\citenamefont {Yang}, \citenamefont {Kimura},\ and\ \citenamefont
		{Otani}(2008)}]{Yang2008}%
	\BibitemOpen
	\bibfield  {author} {\bibinfo {author} {\bibfnamefont {T.}~\bibnamefont
			{Yang}}, \bibinfo {author} {\bibfnamefont {T.}~\bibnamefont {Kimura}}, \ and\
		\bibinfo {author} {\bibfnamefont {Y.}~\bibnamefont {Otani}},\ }\href
	{\doibase 10.1038/nphys1095} {\bibfield  {journal} {\bibinfo  {journal}
			{Nature Physics}\ }\textbf {\bibinfo {volume} {4}},\ \bibinfo {pages} {851}
		(\bibinfo {year} {2008})}\BibitemShut {NoStop}%
	\bibitem [{\citenamefont {Miron}\ \emph {et~al.}(2011)\citenamefont {Miron},
		\citenamefont {Garello}, \citenamefont {Gaudin}, \citenamefont {Zermatten},
		\citenamefont {Costache}, \citenamefont {Auffret}, \citenamefont {Bandiera},
		\citenamefont {Rodmacq}, \citenamefont {Schuhl},\ and\ \citenamefont
		{Gambardella}}]{Miron2011}%
	\BibitemOpen
	\bibfield  {author} {\bibinfo {author} {\bibfnamefont {I.~M.}\ \bibnamefont
			{Miron}}, \bibinfo {author} {\bibfnamefont {K.}~\bibnamefont {Garello}},
		\bibinfo {author} {\bibfnamefont {G.}~\bibnamefont {Gaudin}}, \bibinfo
		{author} {\bibfnamefont {P.-J.}\ \bibnamefont {Zermatten}}, \bibinfo {author}
		{\bibfnamefont {M.~V.}\ \bibnamefont {Costache}}, \bibinfo {author}
		{\bibfnamefont {S.}~\bibnamefont {Auffret}}, \bibinfo {author} {\bibfnamefont
			{S.}~\bibnamefont {Bandiera}}, \bibinfo {author} {\bibfnamefont
			{B.}~\bibnamefont {Rodmacq}}, \bibinfo {author} {\bibfnamefont
			{A.}~\bibnamefont {Schuhl}}, \ and\ \bibinfo {author} {\bibfnamefont
			{P.}~\bibnamefont {Gambardella}},\ }\href {\doibase 10.1038/nature10309}
	{\bibfield  {journal} {\bibinfo  {journal} {Nature}\ }\textbf {\bibinfo
			{volume} {476}},\ \bibinfo {pages} {189} (\bibinfo {year}
		{2011})}\BibitemShut {NoStop}%
	\bibitem [{\citenamefont {Liu}\ \emph {et~al.}(2012)\citenamefont {Liu},
		\citenamefont {Lee}, \citenamefont {Gudmundsen}, \citenamefont {Ralph},\ and\
		\citenamefont {Buhrman}}]{Liu2012}%
	\BibitemOpen
	\bibfield  {author} {\bibinfo {author} {\bibfnamefont {L.}~\bibnamefont
			{Liu}}, \bibinfo {author} {\bibfnamefont {O.~J.}\ \bibnamefont {Lee}},
		\bibinfo {author} {\bibfnamefont {T.~J.}\ \bibnamefont {Gudmundsen}},
		\bibinfo {author} {\bibfnamefont {D.~C.}\ \bibnamefont {Ralph}}, \ and\
		\bibinfo {author} {\bibfnamefont {R.~A.}\ \bibnamefont {Buhrman}},\ }\href
	{\doibase 10.1103/PhysRevLett.109.096602} {\bibfield  {journal} {\bibinfo
			{journal} {Phys. Rev. Lett.}\ }\textbf {\bibinfo {volume} {109}},\ \bibinfo
		{pages} {096602} (\bibinfo {year} {2012})}\BibitemShut {NoStop}%
	\bibitem [{\citenamefont {Weiler}\ \emph {et~al.}(2009)\citenamefont {Weiler},
		\citenamefont {Brandlmaier}, \citenamefont {Gepr{\"a}gs}, \citenamefont
		{Althammer}, \citenamefont {Opel}, \citenamefont {Bihler}, \citenamefont
		{Huebl}, \citenamefont {Brandt}, \citenamefont {Gross},\ and\ \citenamefont
		{Goennenwein}}]{Weiler2009}%
	\BibitemOpen
	\bibfield  {author} {\bibinfo {author} {\bibfnamefont {M.}~\bibnamefont
			{Weiler}}, \bibinfo {author} {\bibfnamefont {A.}~\bibnamefont {Brandlmaier}},
		\bibinfo {author} {\bibfnamefont {S.}~\bibnamefont {Gepr{\"a}gs}}, \bibinfo
		{author} {\bibfnamefont {M.}~\bibnamefont {Althammer}}, \bibinfo {author}
		{\bibfnamefont {M.}~\bibnamefont {Opel}}, \bibinfo {author} {\bibfnamefont
			{C.}~\bibnamefont {Bihler}}, \bibinfo {author} {\bibfnamefont
			{H.}~\bibnamefont {Huebl}}, \bibinfo {author} {\bibfnamefont
			{M.}~\bibnamefont {Brandt}}, \bibinfo {author} {\bibfnamefont
			{R.}~\bibnamefont {Gross}}, \ and\ \bibinfo {author} {\bibfnamefont
			{S.}~\bibnamefont {Goennenwein}},\ }\href@noop {} {\bibfield  {journal}
		{\bibinfo  {journal} {New Journal of Physics}\ }\textbf {\bibinfo {volume}
			{11}},\ \bibinfo {pages} {013021} (\bibinfo {year} {2009})}\BibitemShut
	{NoStop}%
	\bibitem [{\citenamefont {Brandlmaier}\ \emph {et~al.}(2011)\citenamefont
		{Brandlmaier}, \citenamefont {Gepr{\"a}gs}, \citenamefont {Woltersdorf},
		\citenamefont {Gross},\ and\ \citenamefont {Goennenwein}}]{Brandlmaier2011}%
	\BibitemOpen
	\bibfield  {author} {\bibinfo {author} {\bibfnamefont {A.}~\bibnamefont
			{Brandlmaier}}, \bibinfo {author} {\bibfnamefont {S.}~\bibnamefont
			{Gepr{\"a}gs}}, \bibinfo {author} {\bibfnamefont {G.}~\bibnamefont
			{Woltersdorf}}, \bibinfo {author} {\bibfnamefont {R.}~\bibnamefont {Gross}},
		\ and\ \bibinfo {author} {\bibfnamefont {S.}~\bibnamefont {Goennenwein}},\
	}\href@noop {} {\bibfield  {journal} {\bibinfo  {journal} {Journal of Applied
				Physics}\ }\textbf {\bibinfo {volume} {110}},\ \bibinfo {pages} {043913}
		(\bibinfo {year} {2011})}\BibitemShut {NoStop}%
	\bibitem [{\citenamefont {Spaldin}\ and\ \citenamefont
		{Fiebig}(2005)}]{Spaldin2005}%
	\BibitemOpen
	\bibfield  {author} {\bibinfo {author} {\bibfnamefont {N.~A.}\ \bibnamefont
			{Spaldin}}\ and\ \bibinfo {author} {\bibfnamefont {M.}~\bibnamefont
			{Fiebig}},\ }\href@noop {} {\bibfield  {journal} {\bibinfo  {journal}
			{Science}\ }\textbf {\bibinfo {volume} {309}},\ \bibinfo {pages} {391}
		(\bibinfo {year} {2005})}\BibitemShut {NoStop}%
	\bibitem [{\citenamefont {Martin}\ and\ \citenamefont
		{Ramesh}(2012)}]{Martin2012}%
	\BibitemOpen
	\bibfield  {author} {\bibinfo {author} {\bibfnamefont {L.}~\bibnamefont
			{Martin}}\ and\ \bibinfo {author} {\bibfnamefont {R.}~\bibnamefont
			{Ramesh}},\ }\href@noop {} {\bibfield  {journal} {\bibinfo  {journal} {Acta
				Materialia}\ }\textbf {\bibinfo {volume} {60}},\ \bibinfo {pages} {2449}
		(\bibinfo {year} {2012})}\BibitemShut {NoStop}%
	\bibitem [{\citenamefont {Dreher}\ \emph {et~al.}(2012)\citenamefont {Dreher},
		\citenamefont {Weiler}, \citenamefont {Pernpeintner}, \citenamefont {Huebl},
		\citenamefont {Gross}, \citenamefont {Brandt},\ and\ \citenamefont
		{Goennenwein}}]{Dreher2012}%
	\BibitemOpen
	\bibfield  {author} {\bibinfo {author} {\bibfnamefont {L.}~\bibnamefont
			{Dreher}}, \bibinfo {author} {\bibfnamefont {M.}~\bibnamefont {Weiler}},
		\bibinfo {author} {\bibfnamefont {M.}~\bibnamefont {Pernpeintner}}, \bibinfo
		{author} {\bibfnamefont {H.}~\bibnamefont {Huebl}}, \bibinfo {author}
		{\bibfnamefont {R.}~\bibnamefont {Gross}}, \bibinfo {author} {\bibfnamefont
			{M.~S.}\ \bibnamefont {Brandt}}, \ and\ \bibinfo {author} {\bibfnamefont
			{S.~T.~B.}\ \bibnamefont {Goennenwein}},\ }\href {\doibase
		10.1103/PhysRevB.86.134415} {\bibfield  {journal} {\bibinfo  {journal} {Phys.
				Rev. B}\ }\textbf {\bibinfo {volume} {86}},\ \bibinfo {pages} {134415}
		(\bibinfo {year} {2012})}\BibitemShut {NoStop}%
	\bibitem [{\citenamefont {Weiler}\ \emph {et~al.}(2011)\citenamefont {Weiler},
		\citenamefont {Dreher}, \citenamefont {Heeg}, \citenamefont {Huebl},
		\citenamefont {Gross}, \citenamefont {Brandt},\ and\ \citenamefont
		{Goennenwein}}]{Weiler2011}%
	\BibitemOpen
	\bibfield  {author} {\bibinfo {author} {\bibfnamefont {M.}~\bibnamefont
			{Weiler}}, \bibinfo {author} {\bibfnamefont {L.}~\bibnamefont {Dreher}},
		\bibinfo {author} {\bibfnamefont {C.}~\bibnamefont {Heeg}}, \bibinfo {author}
		{\bibfnamefont {H.}~\bibnamefont {Huebl}}, \bibinfo {author} {\bibfnamefont
			{R.}~\bibnamefont {Gross}}, \bibinfo {author} {\bibfnamefont {M.~S.}\
			\bibnamefont {Brandt}}, \ and\ \bibinfo {author} {\bibfnamefont {S.~T.}\
			\bibnamefont {Goennenwein}},\ }\href@noop {} {\bibfield  {journal} {\bibinfo
			{journal} {Physical review letters}\ }\textbf {\bibinfo {volume} {106}},\
		\bibinfo {pages} {117601} (\bibinfo {year} {2011})}\BibitemShut {NoStop}%
	\bibitem [{\citenamefont {Gowtham}\ \emph {et~al.}(2015)\citenamefont
		{Gowtham}, \citenamefont {Moriyama}, \citenamefont {Ralph},\ and\
		\citenamefont {Buhrman}}]{Gowtham2015}%
	\BibitemOpen
	\bibfield  {author} {\bibinfo {author} {\bibfnamefont {P.~G.}\ \bibnamefont
			{Gowtham}}, \bibinfo {author} {\bibfnamefont {T.}~\bibnamefont {Moriyama}},
		\bibinfo {author} {\bibfnamefont {D.~C.}\ \bibnamefont {Ralph}}, \ and\
		\bibinfo {author} {\bibfnamefont {R.~A.}\ \bibnamefont {Buhrman}},\
	}\href@noop {} {\bibfield  {journal} {\bibinfo  {journal} {Journal of Applied
				Physics}\ }\textbf {\bibinfo {volume} {118}},\ \bibinfo {pages} {233910}
		(\bibinfo {year} {2015})}\BibitemShut {NoStop}%
	\bibitem [{\citenamefont {Chang}\ \emph {et~al.}(2018)\citenamefont {Chang},
		\citenamefont {Tamming}, \citenamefont {Broomhall}, \citenamefont
		{Janusonis}, \citenamefont {Fry}, \citenamefont {Tobey},\ and\ \citenamefont
		{Hayward}}]{Chang2018}%
	\BibitemOpen
	\bibfield  {author} {\bibinfo {author} {\bibfnamefont {C.}~\bibnamefont
			{Chang}}, \bibinfo {author} {\bibfnamefont {R.}~\bibnamefont {Tamming}},
		\bibinfo {author} {\bibfnamefont {T.}~\bibnamefont {Broomhall}}, \bibinfo
		{author} {\bibfnamefont {J.}~\bibnamefont {Janusonis}}, \bibinfo {author}
		{\bibfnamefont {P.}~\bibnamefont {Fry}}, \bibinfo {author} {\bibfnamefont
			{R.}~\bibnamefont {Tobey}}, \ and\ \bibinfo {author} {\bibfnamefont
			{T.}~\bibnamefont {Hayward}},\ }\href@noop {} {\bibfield  {journal} {\bibinfo
			{journal} {Physical Review Applied}\ }\textbf {\bibinfo {volume} {10}},\
		\bibinfo {pages} {034068} (\bibinfo {year} {2018})}\BibitemShut {NoStop}%
	\bibitem [{\citenamefont {Grimes}\ \emph {et~al.}(2011)\citenamefont {Grimes},
		\citenamefont {Roy}, \citenamefont {Rani},\ and\ \citenamefont
		{Cai}}]{Grimes2011}%
	\BibitemOpen
	\bibfield  {author} {\bibinfo {author} {\bibfnamefont {C.~A.}\ \bibnamefont
			{Grimes}}, \bibinfo {author} {\bibfnamefont {S.~C.}\ \bibnamefont {Roy}},
		\bibinfo {author} {\bibfnamefont {S.}~\bibnamefont {Rani}}, \ and\ \bibinfo
		{author} {\bibfnamefont {Q.}~\bibnamefont {Cai}},\ }\href@noop {} {\bibfield
		{journal} {\bibinfo  {journal} {Sensors}\ }\textbf {\bibinfo {volume} {11}},\
		\bibinfo {pages} {2809} (\bibinfo {year} {2011})}\BibitemShut {NoStop}%
	\bibitem [{\citenamefont {Schoen}\ \emph {et~al.}(2016)\citenamefont {Schoen},
		\citenamefont {Thonig}, \citenamefont {Schneider}, \citenamefont {Silva},
		\citenamefont {Nembach}, \citenamefont {Eriksson}, \citenamefont {Karis},\
		and\ \citenamefont {Shaw}}]{Schoen2016}%
	\BibitemOpen
	\bibfield  {author} {\bibinfo {author} {\bibfnamefont {M.~A.~W.}\
			\bibnamefont {Schoen}}, \bibinfo {author} {\bibfnamefont {D.}~\bibnamefont
			{Thonig}}, \bibinfo {author} {\bibfnamefont {M.~L.}\ \bibnamefont
			{Schneider}}, \bibinfo {author} {\bibfnamefont {T.~J.}\ \bibnamefont
			{Silva}}, \bibinfo {author} {\bibfnamefont {H.~T.}\ \bibnamefont {Nembach}},
		\bibinfo {author} {\bibfnamefont {O.}~\bibnamefont {Eriksson}}, \bibinfo
		{author} {\bibfnamefont {O.}~\bibnamefont {Karis}}, \ and\ \bibinfo {author}
		{\bibfnamefont {J.~M.}\ \bibnamefont {Shaw}},\ }\href
	{http://www.nature.com/nphys/journal/v12/n9/full/nphys3770.html} {\bibfield
		{journal} {\bibinfo  {journal} {Nat Phys}\ }\textbf {\bibinfo {volume}
			{12}},\ \bibinfo {pages} {839} (\bibinfo {year} {2016})}\BibitemShut
	{NoStop}%
	\bibitem [{\citenamefont {Schoen}\ \emph {et~al.}(2017)\citenamefont {Schoen},
		\citenamefont {Lucassen}, \citenamefont {Nembach}, \citenamefont {Silva},
		\citenamefont {Koopmans}, \citenamefont {Back},\ and\ \citenamefont
		{Shaw}}]{Schoen2017}%
	\BibitemOpen
	\bibfield  {author} {\bibinfo {author} {\bibfnamefont {M.~A.~W.}\
			\bibnamefont {Schoen}}, \bibinfo {author} {\bibfnamefont {J.}~\bibnamefont
			{Lucassen}}, \bibinfo {author} {\bibfnamefont {H.~T.}\ \bibnamefont
			{Nembach}}, \bibinfo {author} {\bibfnamefont {T.~J.}\ \bibnamefont {Silva}},
		\bibinfo {author} {\bibfnamefont {B.}~\bibnamefont {Koopmans}}, \bibinfo
		{author} {\bibfnamefont {C.~H.}\ \bibnamefont {Back}}, \ and\ \bibinfo
		{author} {\bibfnamefont {J.~M.}\ \bibnamefont {Shaw}},\ }\href {\doibase
		10.1103/PhysRevB.95.134410} {\bibfield  {journal} {\bibinfo  {journal} {Phys.
				Rev. B}\ }\textbf {\bibinfo {volume} {95}},\ \bibinfo {pages} {134410}
		(\bibinfo {year} {2017})}\BibitemShut {NoStop}%
	\bibitem [{\citenamefont {Flacke}\ \emph {et~al.}()\citenamefont {Flacke},
		\citenamefont {Liensberger}, \citenamefont {Althammer}, \citenamefont
		{Huebl}, \citenamefont {Geprägs}, \citenamefont {Schultheiss}, \citenamefont
		{Buzdakov}, \citenamefont {Hula}, \citenamefont {Schultheiss}, \citenamefont
		{Edwards}, \citenamefont {Nembach}, \citenamefont {Shaw}, \citenamefont
		{Gross},\ and\ \citenamefont {Weiler}}]{Flacke2019}%
	\BibitemOpen
	\bibfield  {author} {\bibinfo {author} {\bibfnamefont {L.}~\bibnamefont
			{Flacke}}, \bibinfo {author} {\bibfnamefont {L.}~\bibnamefont {Liensberger}},
		\bibinfo {author} {\bibfnamefont {M.}~\bibnamefont {Althammer}}, \bibinfo
		{author} {\bibfnamefont {H.}~\bibnamefont {Huebl}}, \bibinfo {author}
		{\bibfnamefont {S.}~\bibnamefont {Geprägs}}, \bibinfo {author}
		{\bibfnamefont {K.}~\bibnamefont {Schultheiss}}, \bibinfo {author}
		{\bibfnamefont {A.}~\bibnamefont {Buzdakov}}, \bibinfo {author}
		{\bibfnamefont {T.}~\bibnamefont {Hula}}, \bibinfo {author} {\bibfnamefont
			{H.}~\bibnamefont {Schultheiss}}, \bibinfo {author} {\bibfnamefont
			{E.~R.~J.}\ \bibnamefont {Edwards}}, \bibinfo {author} {\bibfnamefont
			{H.~T.}\ \bibnamefont {Nembach}}, \bibinfo {author} {\bibfnamefont {J.~M.}\
			\bibnamefont {Shaw}}, \bibinfo {author} {\bibfnamefont {R.}~\bibnamefont
			{Gross}}, \ and\ \bibinfo {author} {\bibfnamefont {M.}~\bibnamefont
			{Weiler}},\ }\href@noop {} {\ }\Eprint
	{http://arxiv.org/abs/http://arxiv.org/abs/1904.11321v1}
	{http://arxiv.org/abs/1904.11321v1} \BibitemShut {NoStop}%
	\bibitem [{\citenamefont {Collet}\ \emph {et~al.}(2017)\citenamefont {Collet},
		\citenamefont {Gladii}, \citenamefont {Evelt}, \citenamefont {Bessonov},
		\citenamefont {Soumah}, \citenamefont {Bortolotti}, \citenamefont
		{Demokritov}, \citenamefont {Henry}, \citenamefont {Cros}, \citenamefont
		{Bailleul} \emph {et~al.}}]{Collet2017}%
	\BibitemOpen
	\bibfield  {author} {\bibinfo {author} {\bibfnamefont {M.}~\bibnamefont
			{Collet}}, \bibinfo {author} {\bibfnamefont {O.}~\bibnamefont {Gladii}},
		\bibinfo {author} {\bibfnamefont {M.}~\bibnamefont {Evelt}}, \bibinfo
		{author} {\bibfnamefont {V.}~\bibnamefont {Bessonov}}, \bibinfo {author}
		{\bibfnamefont {L.}~\bibnamefont {Soumah}}, \bibinfo {author} {\bibfnamefont
			{P.}~\bibnamefont {Bortolotti}}, \bibinfo {author} {\bibfnamefont
			{S.}~\bibnamefont {Demokritov}}, \bibinfo {author} {\bibfnamefont
			{Y.}~\bibnamefont {Henry}}, \bibinfo {author} {\bibfnamefont
			{V.}~\bibnamefont {Cros}}, \bibinfo {author} {\bibfnamefont {M.}~\bibnamefont
			{Bailleul}},  \emph {et~al.},\ }\href@noop {} {\bibfield  {journal} {\bibinfo
			{journal} {Applied Physics Letters}\ }\textbf {\bibinfo {volume} {110}},\
		\bibinfo {pages} {092408} (\bibinfo {year} {2017})}\BibitemShut {NoStop}%
	\bibitem [{\citenamefont {Hall}(1959)}]{Hall1959}%
	\BibitemOpen
	\bibfield  {author} {\bibinfo {author} {\bibfnamefont {R.~C.}\ \bibnamefont
			{Hall}},\ }\href@noop {} {\bibfield  {journal} {\bibinfo  {journal} {Journal
				of Applied Physics}\ }\textbf {\bibinfo {volume} {30}} (\bibinfo {year}
		{1959})}\BibitemShut {NoStop}%
	\bibitem [{\citenamefont {Hunter}\ \emph {et~al.}(2011)\citenamefont {Hunter},
		\citenamefont {Osborn}, \citenamefont {Wang}, \citenamefont {Kazantseva},
		\citenamefont {Hattrick-Simpers}, \citenamefont {Suchoski}, \citenamefont
		{Takahashi}, \citenamefont {Young}, \citenamefont {Mehta}, \citenamefont
		{Bendersky}, \citenamefont {Lofland}, \citenamefont {Wuttig},\ and\
		\citenamefont {Takeuchi}}]{Hunter2011}%
	\BibitemOpen
	\bibfield  {author} {\bibinfo {author} {\bibfnamefont {D.}~\bibnamefont
			{Hunter}}, \bibinfo {author} {\bibfnamefont {W.}~\bibnamefont {Osborn}},
		\bibinfo {author} {\bibfnamefont {K.}~\bibnamefont {Wang}}, \bibinfo {author}
		{\bibfnamefont {N.}~\bibnamefont {Kazantseva}}, \bibinfo {author}
		{\bibfnamefont {J.}~\bibnamefont {Hattrick-Simpers}}, \bibinfo {author}
		{\bibfnamefont {R.}~\bibnamefont {Suchoski}}, \bibinfo {author}
		{\bibfnamefont {R.}~\bibnamefont {Takahashi}}, \bibinfo {author}
		{\bibfnamefont {M.~L.}\ \bibnamefont {Young}}, \bibinfo {author}
		{\bibfnamefont {A.}~\bibnamefont {Mehta}}, \bibinfo {author} {\bibfnamefont
			{L.~A.}\ \bibnamefont {Bendersky}}, \bibinfo {author} {\bibfnamefont {S.~E.}\
			\bibnamefont {Lofland}}, \bibinfo {author} {\bibfnamefont {M.}~\bibnamefont
			{Wuttig}}, \ and\ \bibinfo {author} {\bibfnamefont {I.}~\bibnamefont
			{Takeuchi}},\ }\href {\doibase 10.1038/ncomms1529} {\bibfield  {journal}
		{\bibinfo  {journal} {Nature Communications}\ }\textbf {\bibinfo {volume}
			{2}},\ \bibinfo {pages} {518} (\bibinfo {year} {2011})}\BibitemShut {NoStop}%
	\bibitem [{\citenamefont {Quandt}\ and\ \citenamefont
		{Ludwig}(1999)}]{Quandt1999}%
	\BibitemOpen
	\bibfield  {author} {\bibinfo {author} {\bibfnamefont {E.}~\bibnamefont
			{Quandt}}\ and\ \bibinfo {author} {\bibfnamefont {A.}~\bibnamefont
			{Ludwig}},\ }\href@noop {} {\bibfield  {journal} {\bibinfo  {journal}
			{Journal of applied physics}\ }\textbf {\bibinfo {volume} {85}},\ \bibinfo
		{pages} {6232} (\bibinfo {year} {1999})}\BibitemShut {NoStop}%
	\bibitem [{\citenamefont {Quandt}\ and\ \citenamefont
		{Ludwig}(2000)}]{Quandt2000}%
	\BibitemOpen
	\bibfield  {author} {\bibinfo {author} {\bibfnamefont {E.}~\bibnamefont
			{Quandt}}\ and\ \bibinfo {author} {\bibfnamefont {A.}~\bibnamefont
			{Ludwig}},\ }\href@noop {} {\bibfield  {journal} {\bibinfo  {journal}
			{Sensors and Actuators A: Physical}\ }\textbf {\bibinfo {volume} {81}},\
		\bibinfo {pages} {275} (\bibinfo {year} {2000})}\BibitemShut {NoStop}%
	\bibitem [{\citenamefont {Quandt}\ \emph {et~al.}(1998)\citenamefont {Quandt},
		\citenamefont {Ludwig}, \citenamefont {Lord},\ and\ \citenamefont
		{Faunce}}]{Quandt1998}%
	\BibitemOpen
	\bibfield  {author} {\bibinfo {author} {\bibfnamefont {E.}~\bibnamefont
			{Quandt}}, \bibinfo {author} {\bibfnamefont {A.}~\bibnamefont {Ludwig}},
		\bibinfo {author} {\bibfnamefont {D.}~\bibnamefont {Lord}}, \ and\ \bibinfo
		{author} {\bibfnamefont {C.}~\bibnamefont {Faunce}},\ }\href@noop {}
	{\bibfield  {journal} {\bibinfo  {journal} {Journal of applied physics}\
		}\textbf {\bibinfo {volume} {83}},\ \bibinfo {pages} {7267} (\bibinfo {year}
		{1998})}\BibitemShut {NoStop}%
	\bibitem [{\citenamefont {Ludwig}\ \emph {et~al.}(2002)\citenamefont {Ludwig},
		\citenamefont {Tewes}, \citenamefont {Glasmachers}, \citenamefont
		{L{\"o}hndorf},\ and\ \citenamefont {Quandt}}]{Ludwig2002}%
	\BibitemOpen
	\bibfield  {author} {\bibinfo {author} {\bibfnamefont {A.}~\bibnamefont
			{Ludwig}}, \bibinfo {author} {\bibfnamefont {M.}~\bibnamefont {Tewes}},
		\bibinfo {author} {\bibfnamefont {S.}~\bibnamefont {Glasmachers}}, \bibinfo
		{author} {\bibfnamefont {M.}~\bibnamefont {L{\"o}hndorf}}, \ and\ \bibinfo
		{author} {\bibfnamefont {E.}~\bibnamefont {Quandt}},\ }\href@noop {}
	{\bibfield  {journal} {\bibinfo  {journal} {Journal of magnetism and magnetic
				materials}\ }\textbf {\bibinfo {volume} {242}},\ \bibinfo {pages} {1126}
		(\bibinfo {year} {2002})}\BibitemShut {NoStop}%
	\bibitem [{\citenamefont {Chopra}\ \emph {et~al.}(2005)\citenamefont {Chopra},
		\citenamefont {Sullivan}, \citenamefont {Ludwig},\ and\ \citenamefont
		{Quandt}}]{Chopra2005}%
	\BibitemOpen
	\bibfield  {author} {\bibinfo {author} {\bibfnamefont {H.~D.}\ \bibnamefont
			{Chopra}}, \bibinfo {author} {\bibfnamefont {M.~R.}\ \bibnamefont
			{Sullivan}}, \bibinfo {author} {\bibfnamefont {A.}~\bibnamefont {Ludwig}}, \
		and\ \bibinfo {author} {\bibfnamefont {E.}~\bibnamefont {Quandt}},\
	}\href@noop {} {\bibfield  {journal} {\bibinfo  {journal} {Physical Review
				B}\ }\textbf {\bibinfo {volume} {72}},\ \bibinfo {pages} {054415} (\bibinfo
		{year} {2005})}\BibitemShut {NoStop}%
	\bibitem [{\citenamefont {Johnson}\ \emph {et~al.}(2004)\citenamefont
		{Johnson}, \citenamefont {Garmestani}, \citenamefont {Chu}, \citenamefont
		{McHenry},\ and\ \citenamefont {Laughlin}}]{Johnson2004}%
	\BibitemOpen
	\bibfield  {author} {\bibinfo {author} {\bibfnamefont {F.}~\bibnamefont
			{Johnson}}, \bibinfo {author} {\bibfnamefont {H.}~\bibnamefont {Garmestani}},
		\bibinfo {author} {\bibfnamefont {S.}~\bibnamefont {Chu}}, \bibinfo {author}
		{\bibfnamefont {M.}~\bibnamefont {McHenry}}, \ and\ \bibinfo {author}
		{\bibfnamefont {D.}~\bibnamefont {Laughlin}},\ }\href@noop {} {\bibfield
		{journal} {\bibinfo  {journal} {IEEE transactions on magnetics}\ }\textbf
		{\bibinfo {volume} {40}},\ \bibinfo {pages} {2697} (\bibinfo {year}
		{2004})}\BibitemShut {NoStop}%
	\bibitem [{\citenamefont {Cooke}, \citenamefont {Gibbs},\ and\ \citenamefont
		{Pettifer}(2001)}]{Cooke2001}%
	\BibitemOpen
	\bibfield  {author} {\bibinfo {author} {\bibfnamefont {M.}~\bibnamefont
			{Cooke}}, \bibinfo {author} {\bibfnamefont {M.}~\bibnamefont {Gibbs}}, \ and\
		\bibinfo {author} {\bibfnamefont {R.}~\bibnamefont {Pettifer}},\ }\href@noop
	{} {\bibfield  {journal} {\bibinfo  {journal} {Journal of magnetism and
				magnetic materials}\ }\textbf {\bibinfo {volume} {237}},\ \bibinfo {pages}
		{175} (\bibinfo {year} {2001})}\BibitemShut {NoStop}%
	\bibitem [{\citenamefont {Cooke}\ \emph {et~al.}(2000)\citenamefont {Cooke},
		\citenamefont {Wang}, \citenamefont {Watts}, \citenamefont {Zuberek},
		\citenamefont {Heydon}, \citenamefont {Rainforth},\ and\ \citenamefont
		{Gehring}}]{Cooke2000}%
	\BibitemOpen
	\bibfield  {author} {\bibinfo {author} {\bibfnamefont {M.}~\bibnamefont
			{Cooke}}, \bibinfo {author} {\bibfnamefont {L.}~\bibnamefont {Wang}},
		\bibinfo {author} {\bibfnamefont {R.}~\bibnamefont {Watts}}, \bibinfo
		{author} {\bibfnamefont {R.}~\bibnamefont {Zuberek}}, \bibinfo {author}
		{\bibfnamefont {G.}~\bibnamefont {Heydon}}, \bibinfo {author} {\bibfnamefont
			{W.}~\bibnamefont {Rainforth}}, \ and\ \bibinfo {author} {\bibfnamefont
			{G.}~\bibnamefont {Gehring}},\ }\href@noop {} {\bibfield  {journal} {\bibinfo
			{journal} {Journal of Physics D: Applied Physics}\ }\textbf {\bibinfo
			{volume} {33}},\ \bibinfo {pages} {1450} (\bibinfo {year}
		{2000})}\BibitemShut {NoStop}%
	\bibitem [{\citenamefont {{\.Z}uberek}\ \emph {et~al.}(2000)\citenamefont
		{{\.Z}uberek}, \citenamefont {Wawro}, \citenamefont {Szymczak}, \citenamefont
		{Wisniewski}, \citenamefont {Paszkowicz},\ and\ \citenamefont
		{Gibbs}}]{Zuberek2000}%
	\BibitemOpen
	\bibfield  {author} {\bibinfo {author} {\bibfnamefont {R.}~\bibnamefont
			{{\.Z}uberek}}, \bibinfo {author} {\bibfnamefont {A.}~\bibnamefont {Wawro}},
		\bibinfo {author} {\bibfnamefont {H.}~\bibnamefont {Szymczak}}, \bibinfo
		{author} {\bibfnamefont {A.}~\bibnamefont {Wisniewski}}, \bibinfo {author}
		{\bibfnamefont {W.}~\bibnamefont {Paszkowicz}}, \ and\ \bibinfo {author}
		{\bibfnamefont {M.}~\bibnamefont {Gibbs}},\ }\href@noop {} {\bibfield
		{journal} {\bibinfo  {journal} {Journal of Magnetism and Magnetic Materials}\
		}\textbf {\bibinfo {volume} {214}},\ \bibinfo {pages} {155} (\bibinfo {year}
		{2000})}\BibitemShut {NoStop}%
	\bibitem [{\citenamefont {Hung}\ \emph {et~al.}(2000)\citenamefont {Hung},
		\citenamefont {Mao}, \citenamefont {Funada}, \citenamefont {Schneider},
		\citenamefont {Miloslavsky}, \citenamefont {Miller}, \citenamefont {Qian},\
		and\ \citenamefont {Tong}}]{Hung2000}%
	\BibitemOpen
	\bibfield  {author} {\bibinfo {author} {\bibfnamefont {C.-Y.}\ \bibnamefont
			{Hung}}, \bibinfo {author} {\bibfnamefont {M.}~\bibnamefont {Mao}}, \bibinfo
		{author} {\bibfnamefont {S.}~\bibnamefont {Funada}}, \bibinfo {author}
		{\bibfnamefont {T.}~\bibnamefont {Schneider}}, \bibinfo {author}
		{\bibfnamefont {L.}~\bibnamefont {Miloslavsky}}, \bibinfo {author}
		{\bibfnamefont {M.}~\bibnamefont {Miller}}, \bibinfo {author} {\bibfnamefont
			{C.}~\bibnamefont {Qian}}, \ and\ \bibinfo {author} {\bibfnamefont
			{H.}~\bibnamefont {Tong}},\ }\href@noop {} {\bibfield  {journal} {\bibinfo
			{journal} {Journal of Applied Physics}\ }\textbf {\bibinfo {volume} {87}},\
		\bibinfo {pages} {6618} (\bibinfo {year} {2000})}\BibitemShut {NoStop}%
	\bibitem [{\citenamefont {Cheng}\ \emph {et~al.}(2018)\citenamefont {Cheng},
		\citenamefont {Lee}, \citenamefont {Brangham}, \citenamefont {White},
		\citenamefont {Ruane}, \citenamefont {Hammel},\ and\ \citenamefont
		{Yang}}]{Cheng2018}%
	\BibitemOpen
	\bibfield  {author} {\bibinfo {author} {\bibfnamefont {Y.}~\bibnamefont
			{Cheng}}, \bibinfo {author} {\bibfnamefont {A.~J.}\ \bibnamefont {Lee}},
		\bibinfo {author} {\bibfnamefont {J.~T.}\ \bibnamefont {Brangham}}, \bibinfo
		{author} {\bibfnamefont {S.~P.}\ \bibnamefont {White}}, \bibinfo {author}
		{\bibfnamefont {W.~T.}\ \bibnamefont {Ruane}}, \bibinfo {author}
		{\bibfnamefont {P.~C.}\ \bibnamefont {Hammel}}, \ and\ \bibinfo {author}
		{\bibfnamefont {F.}~\bibnamefont {Yang}},\ }\href@noop {} {\bibfield
		{journal} {\bibinfo  {journal} {Applied Physics Letters}\ }\textbf {\bibinfo
			{volume} {113}},\ \bibinfo {pages} {262403} (\bibinfo {year}
		{2018})}\BibitemShut {NoStop}%
	\bibitem [{\citenamefont {Pernpeintner}\ \emph {et~al.}(2016)\citenamefont
		{Pernpeintner}, \citenamefont {Holländer}, \citenamefont {Seitner},
		\citenamefont {Weig}, \citenamefont {Gross}, \citenamefont {Goennenwein},\
		and\ \citenamefont {Huebl}}]{Pernpeintner2016}%
	\BibitemOpen
	\bibfield  {author} {\bibinfo {author} {\bibfnamefont {M.}~\bibnamefont
			{Pernpeintner}}, \bibinfo {author} {\bibfnamefont {R.~B.}\ \bibnamefont
			{Holländer}}, \bibinfo {author} {\bibfnamefont {M.~J.}\ \bibnamefont
			{Seitner}}, \bibinfo {author} {\bibfnamefont {E.~M.}\ \bibnamefont {Weig}},
		\bibinfo {author} {\bibfnamefont {R.}~\bibnamefont {Gross}}, \bibinfo
		{author} {\bibfnamefont {S.~T.~B.}\ \bibnamefont {Goennenwein}}, \ and\
		\bibinfo {author} {\bibfnamefont {H.}~\bibnamefont {Huebl}},\ }\href@noop {}
	{\bibfield  {journal} {\bibinfo  {journal} {Journal of Applied Physics}\
		}\textbf {\bibinfo {volume} {119}},\ \bibinfo {pages} {093901} (\bibinfo
		{year} {2016})}\BibitemShut {NoStop}%
	\bibitem [{\citenamefont {Timoshenko}, \citenamefont {Weaver},\ and\
		\citenamefont {Young}(h ed)}]{Timoshenko2008}%
	\BibitemOpen
	\bibfield  {author} {\bibinfo {author} {\bibfnamefont {S.}~\bibnamefont
			{Timoshenko}}, \bibinfo {author} {\bibfnamefont {W.}~\bibnamefont {Weaver}},
		\ and\ \bibinfo {author} {\bibfnamefont {D.}~\bibnamefont {Young}},\
	}\href@noop {} {\emph {\bibinfo {title} {Vibration Problems in
				Engeneering}}}\ (\bibinfo  {publisher} {John Wiley and Sons: New York},\
	\bibinfo {year} {1990 5th ed.})\BibitemShut {NoStop}%
	\bibitem [{\citenamefont {Verbridge}\ \emph {et~al.}(2006)\citenamefont
		{Verbridge}, \citenamefont {Parpia}, \citenamefont {Reichenbach},
		\citenamefont {Bellan},\ and\ \citenamefont {Craighead}}]{Verbridge2006}%
	\BibitemOpen
	\bibfield  {author} {\bibinfo {author} {\bibfnamefont {S.~S.}\ \bibnamefont
			{Verbridge}}, \bibinfo {author} {\bibfnamefont {J.~M.}\ \bibnamefont
			{Parpia}}, \bibinfo {author} {\bibfnamefont {R.~B.}\ \bibnamefont
			{Reichenbach}}, \bibinfo {author} {\bibfnamefont {L.~M.}\ \bibnamefont
			{Bellan}}, \ and\ \bibinfo {author} {\bibfnamefont {H.~G.}\ \bibnamefont
			{Craighead}},\ }\href {\doibase 10.1063/1.2204829} {\bibfield  {journal}
		{\bibinfo  {journal} {Journal of Applied Physics}\ }\textbf {\bibinfo
			{volume} {99}},\ \bibinfo {pages} {124304} (\bibinfo {year}
		{2006})}\BibitemShut {NoStop}%
	\bibitem [{\citenamefont {Gross}\ and\ \citenamefont {Marx}(2012)}]{GrossMarx}%
	\BibitemOpen
	\bibfield  {author} {\bibinfo {author} {\bibfnamefont {R.}~\bibnamefont
			{Gross}}\ and\ \bibinfo {author} {\bibfnamefont {A.}~\bibnamefont {Marx}},\
	}\href@noop {} {\emph {\bibinfo {title} {Festkörperphysik}}}\ (\bibinfo
	{publisher} {De Gruyter Oldenbourg},\ \bibinfo {year} {2012})\BibitemShut
	{NoStop}%
	\bibitem [{\citenamefont {Chikazumi}(1997)}]{Chikazumi}%
	\BibitemOpen
	\bibfield  {author} {\bibinfo {author} {\bibfnamefont {S.}~\bibnamefont
			{Chikazumi}},\ }\href@noop {} {\emph {\bibinfo {title} {Physics of
				Ferromagnetism (International Series of Monographs on Physics)}}}\ (\bibinfo
	{publisher} {Clarendon Press},\ \bibinfo {year} {1997})\BibitemShut {NoStop}%
	\bibitem [{\citenamefont {Seitner}, \citenamefont {Gajo},\ and\ \citenamefont
		{Weig}(2014)}]{Seitner2014}%
	\BibitemOpen
	\bibfield  {author} {\bibinfo {author} {\bibfnamefont {M.~J.}\ \bibnamefont
			{Seitner}}, \bibinfo {author} {\bibfnamefont {K.}~\bibnamefont {Gajo}}, \
		and\ \bibinfo {author} {\bibfnamefont {E.~M.}\ \bibnamefont {Weig}},\
	}\href@noop {} {\bibfield  {journal} {\bibinfo  {journal} {Applied Physics
				Letters}\ }\textbf {\bibinfo {volume} {105}},\ \bibinfo {pages} {213101}
		(\bibinfo {year} {2014})}\BibitemShut {NoStop}%
	\bibitem [{\citenamefont {Hoehne}\ \emph {et~al.}(2010)\citenamefont {Hoehne},
		\citenamefont {Pashkin}, \citenamefont {Astafiev}, \citenamefont {Faoro},
		\citenamefont {Ioffe}, \citenamefont {Nakamura},\ and\ \citenamefont
		{Tsai}}]{Hoehne2010}%
	\BibitemOpen
	\bibfield  {author} {\bibinfo {author} {\bibfnamefont {F.}~\bibnamefont
			{Hoehne}}, \bibinfo {author} {\bibfnamefont {Y.~A.}\ \bibnamefont {Pashkin}},
		\bibinfo {author} {\bibfnamefont {O.}~\bibnamefont {Astafiev}}, \bibinfo
		{author} {\bibfnamefont {L.}~\bibnamefont {Faoro}}, \bibinfo {author}
		{\bibfnamefont {L.}~\bibnamefont {Ioffe}}, \bibinfo {author} {\bibfnamefont
			{Y.}~\bibnamefont {Nakamura}}, \ and\ \bibinfo {author} {\bibfnamefont
			{J.}~\bibnamefont {Tsai}},\ }\href@noop {} {\bibfield  {journal} {\bibinfo
			{journal} {Physical Review B}\ }\textbf {\bibinfo {volume} {81}},\ \bibinfo
		{pages} {184112} (\bibinfo {year} {2010})}\BibitemShut {NoStop}%
	\bibitem [{\citenamefont {Cardarelli}(2018)}]{Cardarelli2018}%
	\BibitemOpen
	\bibfield  {author} {\bibinfo {author} {\bibfnamefont {F.}~\bibnamefont
			{Cardarelli}},\ }in\ \href {\doibase 10.1007/978-3-319-38925-7_2} {\emph
		{\bibinfo {booktitle} {Materials Handbook: A Concise Desktop Reference}}}\
	(\bibinfo  {publisher} {Springer International Publishing},\ \bibinfo {year}
	{2018})\ pp.\ \bibinfo {pages} {101--248}\BibitemShut {NoStop}%
	\bibitem [{\citenamefont {Edwards}, \citenamefont {Nembach},\ and\
		\citenamefont {Shaw}(2019)}]{Edwards2019}%
	\BibitemOpen
	\bibfield  {author} {\bibinfo {author} {\bibfnamefont {E.~R.}\ \bibnamefont
			{Edwards}}, \bibinfo {author} {\bibfnamefont {H.~T.}\ \bibnamefont
			{Nembach}}, \ and\ \bibinfo {author} {\bibfnamefont {J.~M.}\ \bibnamefont
			{Shaw}},\ }\href@noop {} {\bibfield  {journal} {\bibinfo  {journal} {Physical
				Review Applied}\ }\textbf {\bibinfo {volume} {11}},\ \bibinfo {pages}
		{054036} (\bibinfo {year} {2019})}\BibitemShut {NoStop}%
\end{thebibliography}
%

\end{document}